\documentstyle[preprint,prc,aps,psfig]{revtex}
\everymath={\displaystyle}
\begin{document}
%\draf
%\twocolumn[
%\vsize=1.5cm
\title{\bf On the exact solutions of the Lipkin-Meshkov-Glick model}

\author{N. Debergh\footnote{e-mail address: Nathalie.Debergh@ulg.ac.be
; chercheur I.I.S.N.}
and Fl. Stancu\thanks{ e-mail address: fstancu@ulg.ac.be 
%\newline
%address until May 31, 2000: ECT*, Strada delle Tabarelle,
%286, I-38050 Villazzano, Trento, Italy,
%e-mail address:
~~or ~~stancu@ect.it (until May 31, 2001)}$^*$}

\address{$^*$University of Li\`ege, Institute of Physics B5, Sart Tilman, B-4000
Li\`ege 1, Belgium}
\address{$^\dagger$ECT*, Strada delle Tabarelle, 286, I-38050 Villazzano, Trento, Italy}

%\vspace{1.5cm}
\date{\today}
\maketitle

\begin{abstract}
\baselineskip=0.50cm
We present the many-particle Hamiltonian model of Lipkin, Meshkov and Glick
in the context of deformed polynomial algebras and show that its 
exact solutions can be easily and naturally obtained within this formalism.
The Hamiltonian matrix of each $j$ multiplet can be split into two
submatrices associated to two distinct irreps of the deformed algebra.
Their invariant subspaces correspond to even and odd numbers of 
particle-hole excitations.
\end{abstract}
%%%%%%%%%%%%%%%%%%%%%%%%%%%%%%%%%%%%%%%%%%%%%%%%%%%%%%%%%%%%%%%%%%%%%%%%%%%%%%%
\section{Introduction}
In the sixties much interest has been devoted 
to formalisms for treating multiparticle systems and
the quality of the approximations involved. To test the validity
of the approximations, quasi-exactly solvable models 
have been proposed (for a definition of a quasi-exactly solvable model
see e.g. Ref. \cite{TURBINER}).
The comparison between the exact solutions and an approximation could
give a clear estimate of the quality of the approximation,
which could further be applied to more complicated Hamiltonians.
Among them of particular interest is the model of Lipkin, Meshkov and
Glick (LMG) \cite{LMG1}.  
Although simple enough to be solved exactly,
in some cases the model is not trivial. Few analytic solutions
(for a number of particles up to 8) 
have been provided by LMG.  Numerical solutions were also given 
for a larger number of particles in the case where the 
total angular momentum reaches its maximum value. 
Here we study the exact solutions of the LMG model in the framework 
of deformed algebras. We first show that the LMG model corresponds
to a deformed algebra of polynomial type and then we search for
possible solutions associated with the representations 
of the corresponding deformed algebra.
We show that the polynomial algebra introduces a new symmetry
in the system, not known before, and this splits into two submatrices
any Hamiltonian matrix to be diagonalized for a specific value of 
the angular momentum. 
Moreover, it introduces a new quantum number which 
naturally distinguishes between an even and an odd number 
of particles.\par

In the next section we recall the LMG model.
In Sec. III we shortly introduce deformed algebras and describe 
the way one can use them to find exact solutions of the LMG model
for an arbitrary given number of particles and for any specific value
$j$ of the angular momentum. In Sec. IV we presents algebraic and 
numerical solutions for a system with an odd ($N$ = 7) and an even
($N$ = 8) number of particles. A general description of some 
supplementary solutions, 
inherent to the deformed algebra, is given in Sec. V.
The last section is devoted to a summary.
%%%%%%%%%%%%%%%%%%%%%%%%%%%%%%%%%%%%%%%%%%%%%%%%%%%%%%%%%%%%%
\section{The LMG model}
As mentioned in the introduction the model of Lipkin, Meshkov and
Glick is a quasi-exactly soluble model developed for treating
many particle systems.\par
The general method for constructing soluble models is based on the
incorporation of some symmetries of the system which give additional
integrals of motion and therefore reduce the size of the Hamiltonian
matrix to be diagonalized.
The Hamiltonian of a many-particle system interacting via a two-body
force is a sum of  linear and quadratic terms in bilinear products
of creation and annihilation operators related to the quantum states
of these particles.
One starts from the observation that
bilinear products of creation and annihilation operators can be considered
as elements of a Lie algebra, here related   
in particular to the SU(2) group of
the so-called quasi-spin. Lipkin,
Meshkov and Glick construct a two $N$-fold degenerate level model
where $N$ is the number of fermions in the system. 
The two levels are separated by an energy $\epsilon$. 
Here we discuss the simplified version of the LMG model where the
interaction contains only terms which mixes configurations.
In the following we use the notations of Ref. \cite{KM}.
Accordingly we introduce fermion operators $\beta^+_m, \beta_m$ 
that create and annihilate holes in the
lower level and $\alpha^+_m, \alpha_m$ ($m = 1,2,...,N$)
that create and annihilate particles in the upper level.
These operators satisfy the anticommutation relations 
\begin{equation}\label{COMM}
\{{\alpha_m~,~\alpha^+_{m'}}\} = \{{\beta_m~,~\beta^+_{m'}}\} 
= \delta_{m m'}~,
\end{equation}
as well as the commutation relations
\begin{equation}\label{COMMREL}
[\alpha_m~,~\beta_{m'}] = [\alpha_m~,~\beta^+_{m'}] =
[\alpha^+_m~,~\beta_{m'}] = [\alpha^+_m~,~\beta^+_{m'}] = 0~.
\end{equation}
The ground state $ |0 \rangle$
is defined by
\begin{equation}\label{GS}
\alpha_m~  |0 \rangle = \beta_m~ |0 \rangle = 0.
\end{equation}
Then the bilinear products
\begin{equation}\label{J0}
j_0 = -\frac{1}{2}~N + \frac{1}{2}~\sum_{m=1}^{N} 
~(\alpha^+_m~ \alpha_m + \beta^+_m~ \beta_m)~,
\end{equation}
\begin{equation}\label{J+}
j_+ = \sum_{m=1}^{N}~\alpha^+_m~ \beta^+_m~,
\end{equation}
\begin{equation}\label{J-}
j_- = \sum_{m=1}^{N}~\alpha_m~ \beta_m~,
\end{equation}
form an su(2) algebra.
The Hamiltonian under study can be written in 
terms of these generators as
\begin{equation}\label{HAM}
H_{LMG} = \epsilon~j_0 + V~(j^2_+ + j^2_-)~,
\end{equation} 
where $\epsilon$ is the separation energy between the two levels,
as introduced above, and $V$ is the interaction strength. For later 
purposes it is convenient to introduce the strength parameter
$\delta$ instead of $V$. They are related by
\begin{equation}\label{V}
V = \frac{\delta~ \epsilon}{2~N}
\end{equation}

The invariant operator of the su(2) algebra
\begin{equation}\label{SU2}
j^2 = 1/2(j_+j_- + j_-j_+) + j^2_0
\end{equation}
commutes with the Hamiltonian and provides a constant of motion.
Thus the Hamiltonian matrix breaks up into submatrices each associated with a 
different value of $j$ and of order $2j+1$. 
Each state in a $j$ multiplet has a different number of excited particle-hole
pairs. The interaction (\ref{HAM}) mixes the states within the same $j$ 
multiplet but cannot mix states having different eigenvalues of $j^2$.
It can only excite or de-excite two particle-hole
pairs or in other words it can only change the eigenvalue of $j_0$ by
two units. The eigenstates of $H_{LMG}$ have therefore an important property,
namely their structure is compatible with Hartree Fock solutions, 
so one can study the validity
of the Random Phase Approximation, often used in the treatment of
a system of fermions, against the exact solution. From
the definitions (\ref{COMM})-(\ref{J0}) it follows that
the eigenvalues of $j_0$ are given by half the difference between the number of
particles in the upper level and the number of particles in the
lower level. Then the maximum eigenvalue of $j_0$ and of $j$ is ${N}/2$.
Thus LMG conclude that the largest matrix to be
diagonalized is of dimension $2j+1$ = $N$ + 1. 

Then fixing the number of particles $N$, LMG diagonalize the largest Hamiltonian
matrix associated with $j = N/2$ for 
several cases. For $N$ = 2, 3, 4, 6 and 8 analytical solutions
are provided. In addition, the eigenvalues of
the multiplet $j = N/2$ are found numerically for $N$ = 14, 30 and 50.
Here by using the polynomial 
algebra technique we extend the study of the LMG model to its
entire spectrum.
We show that for a given number of particles there are two types of 
states:
1) states with $j = N/2$,
for which the interaction entirely lifts the degeneracy. One of
these states corresponds to the lowest eigenvalue. These are the
states analysed by LMG and they belong to the largest matrix
to be diagonalized of dimension
$N$ + 1 ;
2) states with $j < N/2$, for which the eigenvalues
are identical with those of  a system with $N$ - 2, $N$ - 4,...
The only difference is that these states are degenerate in a system
with $N$ particles but not degenerate in a system with
$N$ -2, $N$ - 4, ...particles, because there 
they are states of type 1. Thus finding the eigenstates of
a system with $N$ particles reduces to the diagonalization of the
largest matrix once the states of the system with $N$ - 2
particles is known.

In the context of the deformed
polynomial algebra we show that the largest matrix associated to a 
given $N$ can be split into two submatrices
of dimensions $N/2$ + 1 and  $N/2$ for $N$ even and two 
submatrices, both  of dimensions $(N + 1)/2$ for $N$ odd. These
submatrices correspond
to specific values of the Casimir operator of the deformed
algebra. The same statement holds for the largest matrix of a
system of $N$ - 2 particles, and so on. 
Alternatively, for any j multiplet the corresponding Hamiltonian matrix
can be split into two submatrices irrespective of the number of
particles.
These findings are illustrated in detail for the cases of $N = 7$ and
$N$ = 8 particles. 

The splitting of a j multiplet into two submatrices is entirely
consistent with the property of the LMG interaction (\ref{HAM}) that it
can excite or de-excite only two particle-hole pairs. Accordingly,
for $N$ even, on the one hand the states with $0,2,...,N-2,N$ particle-hole 
form an $N/2+1$ dimension invariant subspace and on the other hand the states
with $1,3,...,N-1$ particle-hole excitations form another $N/2$ dimensional
invariant subspace. For $N$ odd the states with $0,2,...,N-1$ particle-hole
excitations and the states with $1,3,...,N$ particle-hole excitations
form two distinct invariant subspaces both of dimension $(N+1)/2$.
The deformed polynomial algebra provides a "quantum number" denoted 
here by {\it c} to
distinguish between the eigenvalue of (\ref{HAM}) for $N$ even
and $N$ odd. For $N$ even one has $c = 0$ and for $N$ odd $c =\pm 1/4$.  
 
Moreover the polynomial algebra technique leads to new representations
corresponding to new eigenvalues. Some of these are appropriate 
to a generalized type of LMG model, some others are meaningless
(see Sec. V).

As far as physics is concerned one should mention that the LMG model
posesses states of collective excitations related for example in
nuclear physics to giant resonances. For this reason
Lipkin Meshkov and Glick studied cases with a  number of particles and
interaction strength relevant
to the treatment of nuclei by the Random Phase Approximation.
In further studies \cite{LMG2,LMG3} they tested
the method of linearizing the equations of
motion and the diagram summation approximations
against exact solutions of their model given in \cite{LMG1}.
Recently a more general Hamiltonian which can test two types of
elementary excitations instead of one has been proposed by Lipkin \cite{L99}.  
%%%%%%%%%%%%%%%%%%%%%%%%%%%%%%%%%%%%%%%%%%%%%%%%%%%%%%%%%%%%%%%%%
\section{Deformed polynomial algebra}
Here we present the LMG model in the context of the deformed
algebra. In doing so we follow the general method based on the
polynomial deformations of the Lie algebra $sl(2,R)$ as
developed in Ref.\cite{DEBERGH}.

We start by noting that the Hamiltonian (\ref{HAM}) is a particular
realization of the more general Hamiltonian
\begin{equation}\label{HAMDEF}
H = \epsilon~[2 J_0 + \delta (J_+ + J_-)]~,
\end{equation}
with the parameter $\epsilon$ as defined above, $\delta$ defined
by Eq. (\ref{V}) and
\begin{equation}\label{GEN} 
J_0 = \frac{1}{2}~j_0, ~~~~~J_{\pm} = \frac{1}{2 N}~j_{\pm}^2~.
\end{equation} 
One can show that the operators (\ref{GEN}) satisfy the following algebra
\begin{equation}\label{DF1}
[J_0,J_{\pm}] = \pm J_{\pm}~,
\end{equation}
\begin{equation}\label{DF2} 
[J_+,J_-] = -\frac{16}{N^2}~J^3_0 + \frac{2}{N^2}
(2 j^2 + 2 j - 1)J_0~,
\end{equation}
where §j§ is an eigenvalue of (\ref{SU2}). 
%The relations (\ref{DF1})
%and (\ref{DF2}) can be recovered from the $sl(2)$ algebra by taking
%\begin{equation} 
%J_0 = \frac{1}{2}~j_0, ~~~~~J_{\pm} = \frac{1}{2 \Omega}~j_{\pm}~.
%\end{equation} 
These relations define a particular case of a deformed polynomial algebra as
studied in Ref. \cite{DEBERGH} with the polynom in 
$J_0$ in the right hand side of (\ref{DF2}) having the coefficients
(see Appendix)
\begin{equation}\label{PARAM}
\alpha = -\frac{16}{N^2}~,~~~\beta = 0,
~~~\gamma = \frac{2}{N^2}(2 j^2 + 2 j - 1),~~~\Delta = 0~.
\end{equation}
The Casimir operator of this algebra is given by
\begin{equation}\label{CASIMIR}
C = J_+~J_- -\frac{4}{N^2}J^4_0 + \frac{8}{N^2}J^3_0
+ \frac{2 j^2 + 2 j -5}{N^2}J^2_0 - 
\frac{2 j^2 + 2 j - 1}{N^2}J_0
\end{equation}
%The part of Ref. \cite{DEBERGH} relevant for this study is summarized
%in Appendix A.
The algebra (\ref{DF1})-(\ref{DF2}) has two types of
representations relevant for our discussion. They are
labelled by $q = 1$ and $q = 2$ respectively.
More precisely, the $q = 1$ representations
are defined by the equations
\begin{eqnarray}\label{QEQUAL1}
J_0~ |JM \rangle & = & (M + c)~ |JM \rangle~,\nonumber \\
J_+~ |JM \rangle & = & f(M)~ |J,M + 1 \rangle~,\nonumber \\
J_-~ |JM \rangle & = & g(M)~ |J,M - 1 \rangle~,
\end{eqnarray}
with $M = -J,...,J, ~J = 0, \frac{1}{2},...~~~ ,~ c~ \epsilon~ \Re $ and
\begin{eqnarray}\label{FG}
f(M - 1)~g(M) & = & \frac{1}{N^2}(J-M+1)(J+M) \nonumber \\
 & \times & [2 j^2 + 2j - 1 - 4 J^2 - 4J - 4 M^2 + 4M + 8(1-2M)~c - 24 c^2]~,
\end{eqnarray}
where the real number $c$, constrained by Eq. (\ref{Q1}) (see Appendix),
can take three distinct
values given by 
\begin{equation}
c = 0~,
\end{equation}
or
\begin{equation}
c = \pm~ [ \frac{1}{4} j(j + 1) - \frac{1}{8} - J(J + 1) ]^{1/2}~.
\end{equation}

The $q = 2$ representations are defined in an invariant subspace
satisfying
\begin{eqnarray}\label{JINTEGER}
J_0~|J^{'} M^{'} \rangle & = & \frac{M^{'}}{2}~ |J^{'} M^{'} \rangle~,\nonumber \\
J_+~ |J^{'} M^{'} \rangle & = & f^{'}(M^{'})~ |J^{'},M^{'} + 2 \rangle~,\nonumber \\
J_-~ |J^{'} M^{'} \rangle & = & g^{'}(M^{'})~ |J^{'},M^{'} - 2 \rangle~,
\end{eqnarray} 
where $J^{'} = 0,1,2,...$~~and
\begin{eqnarray}
f^{'}(M^{'} - 2)g{'}(M^{'}) & = & \frac{1}{4 N^2}(J^{'} - M^{'} + 2)(J^{'} + M^{'})
\nonumber \\
 & \times & (2 j^2 + 2 j - 1 - J^{'2} - 2 J^{'} - M^{'2} + 2 M^{'})~,
\end{eqnarray}
if $M^{'} = - J^{'},...,J^{'} - 2, J^{'}  $ and
\begin{eqnarray}
f^{'}(M^{'} - 2)g{'}(M^{'}) & = & \frac{1}{4 N^2}(J^{'} - M^{'} + 1)(J^{'} + M^{'} - 1)
\nonumber \\
 & \times  & (2 j^2 + 2 j - J^{'2} - M^{'2} + 2 M^{'})~,
\end{eqnarray}  
if $M^{'} = - J^{'} + 1,...,J^{'} - 3, J^{'} - 1  $.  

The cases $J^{'} =
\frac{1}{2}, \frac{3}{2},...$~~ are particular in the sense that $J{'}$
must be equal to $j$. The relations satisfied by the basis vectors 
$|jm \rangle$ are 
\begin{eqnarray}\label{JHALF}
J_0~ |jm \rangle & = & \frac{m}{2} |jm \rangle~, \nonumber \\
J_+~ |jm \rangle & = & f^{'}(m)~ |j,m + 2 \rangle~,\nonumber \\
J_-~ |jm \rangle & = & g^{'}(m)~ |j,m - 2 \rangle~,
\end{eqnarray}
with
\begin{equation}
f^{'}(m - 2)g^{'}(m) = \frac{1}{4 N^2}(j+m)(j+m-1)(j-m+1)(j-m+2)~.
\end{equation}

The Hamiltonian (\ref{HAM}) 
can be associated to the representation §q = 2§ since in this case the 
invariant subspace is spanned by the vectors $|jm \rangle$ on which
the deformed generators act as follows 
\begin{equation}
J_0 |jm \rangle = \frac{m}{2} |jm \rangle~,
\end{equation}
\begin{equation}
J_+ |jm \rangle = \frac{1}{2N} \sqrt{(j-m-1)(j-m)(j+m+1)(j+m+2)} 
~|j,m+2 \rangle~,
\end{equation}
\begin{equation}
J_- |jm \rangle = \frac{1}{2N} \sqrt{(j+m-1)(j+m)(j-m+1)(j-m+2)} 
~|j,m-2 \rangle~.
\end{equation} 
One can see that these relations can be recovered from the equations
(\ref{JHALF}) if $j$ is a half integer and from (\ref{JINTEGER}) 
but with $J^{'} = j$ if $j$ is an integer.

If one now calculates the eigenvalues of the Casimir operator (\ref{CASIMIR})
for the representation $q = 2$ and $J^{'} = n$, i.e. an integer, one gets
\begin{equation}
{\langle C \rangle}_{J^{'}=n,q=2} =
\left(
\begin{array}[]{lc}
\frac{1}{2 N^2}~ n(n+2)[j(j+1)-\frac{1}{2} (n+1)^2]~~;&
~~~~~M^{'}=n,n-2,...,-n \\
\frac{1}{2 N^2}~ (n-1)(n-2)[j(j+1)- \frac{1}{2} n^2]~~;&
~~~~~M^{'}=n-1,n-3,...,-n+1
\end{array}
\right)
\end{equation} 
One can therefore see that $C$ has two distinct eigenvalues in the space
spanned by $ | J^{'} M^{'} > $. 
This shows that the representation
$q = 2$ is reducible. One can easily prove that it can be split
into the direct sum
\begin{equation}\label{NEVEN}
{(J^{'}=n, q=2)}_{c=0}~~~ =~~~ {(J=\frac{n}{2}, q=1)}_{c=0} \bigoplus 
{(J=\frac{n-1}{2},q=1)}_{c=0}
\end{equation}
i.e. the $q=2$ representation can be decomposed into two $q=1$ representations
and this takes place for $c=0$ only.

A similar decomposition also holds for half integer $j$.
In this case one has 
\begin{equation}\label{NODD}
{(j=n+\frac{1}{2},q=2)}_{c=0}~~~ =~~~ {(J=\frac{n}{2}, q=1)}_{c=1/4} 
\bigoplus
{(J=\frac{n}{2}, q=1)}_{c=-1/4}
\end{equation}
and the eigenvalue of the Casimir operator is the same for $q=1$ and
$q=2$
\begin{equation}
{\langle C \rangle}_{j=n+1/2} =
\frac{1}{4 N^2}~j(j-1)(j+1)(j+2)~.
\end{equation} 
We can then conclude that the Hamiltonian matrix of each $j$ multiplet
can be split into two submatrices. Examples are shown in the next section.

>From now on we are concerned with $q = 1$ representations only.
Then in the invariant subspace defined by Eqs. (\ref{QEQUAL1}) we obtain 
the following Hamiltonian matrix to be diagonalized  
\begin{equation}
\langle H \rangle =
\left(
\begin{array}[]{ccccccc}
2J+2c & \delta f(J-1) & ~~~0 & 0 &. &~~~ . & 0 \\
\delta g(J) & 2J-2+2c & ~~~\delta f(J-2)~~~ & 0 & .&~~~ .&0 \\
0 & \delta g(J-1)~~~ &~~~ 2J-4+2c & \delta f(J-3) & . &~~~ . & 0 \\
0 & 0 & ~~~\delta g(J-2)~~~ & 2J-6+2c~~~ & . &~~~ . & 0 \\
. & . & ~~~. & . & . &~~~ . & .\\
. & . & ~~~. & . & . &~~~ -2J+2+2c & \delta f(-J)\\
0 & 0 & ~~~0 & 0 & . &~~~ \delta g(-J+1) &-2J+2c    
\end{array}
\right)
\end{equation}
The diagonalization amounts to solving the secular equation
\begin{equation}\label{SEC}
det |{\langle H \rangle }_{ij} - E \delta_{ij} | = 0
\end{equation}
%%%%%%%%%%%%%%%%%%%%%%%%%%%%%%%%%%%%%%%%%%%%%%%%%%%%%%%%%%%%%%%%%%%
\section{The cases  $N$ = 7 and 8 particles}
In this section we give analytic and numerical results for the eigenvalues 
of (\ref{HAM}) obtained by solving the eigenvalue equation 
(\ref{SEC}) for 7 and 8 particles. For a larger number of particles
we checked that we perfectly agree with the numerical values of
Ref. \cite{LMG1}.
%%%%%%%%%%%%%%%%%%%%%%%%%%%%%%%%%%%%%%%%%%%%%%%%%%%%%%%%%%%%%%%%
\subsection{$N = 7$}
For $N=7$ there are $2^7$=128 states. The largest Hamiltonian matrix  
corresponds to $j=7/2$. According to the relation (\ref{NODD})
where we have to take $n=(N-1)/2$ this matrix splits into two equal
matrices, both having $J=(N-1)/4$. The same procedure applies to the
$j-1$ multiplet which is the largest multiplet of $N=5$ particles, and
so on. In Table 1 we give the possible $j$ multiplets, their multiplicities
$m_j$ and the corresponding values of $J$. Analytic forms
of the eigenvalues can be easily obtained only for $j=1/2$ (trivial case)
and $j=3/2$. For $j=5/2$ they are numerically obtained from the
secular equation (\ref{SEC}) which in this case becomes
\begin{equation}
E^3 - 6 c E^2 -(\frac{13}{4}+ \frac{4}{7} {\delta}^2) E
+ \frac{15}{2} c + \frac{120}{49} c {\delta}^2 = 0
\end{equation}
with $c=\pm 1/4$.
For $j=7/2$ the secular equation (\ref{SEC}) leads to
\begin{equation}
E^4 - 8 c E^3 -(\frac{17}{2}+ \frac{18}{7} {\delta}^2) E^2 
+(38+ \frac{888}{49} {\delta}^2) c E 
+\frac{105}{16} + \frac{75}{14} {\delta}^2 + \frac{135}{343} {\delta}^4 = 0
\end{equation}
with $c=\pm 1/4$
which is also solved numerically. The dependence of the eigenvalues on the
parameter $\delta$ is exhibited in Fig. 1.
 
%%%%%%%%%%%%%%%%%%%%%%%%%%%%%%%%%%%%%%%%%%%%%%%%%%%%%%%%%%%%%%%%%%
\subsection{$N = 8$}
For $N=8$ there are 256 states. This case
is more fortunate because one can push the
analytic calculation further than for $N=7$. 
The largest Hamiltonian matrix  
corresponds to $j = N/2 = 4$. According to (\ref{NEVEN}) where we have
to take $n = N/2$ this matrix is split into two submatrices
one corresponding to $J = N/4 = 2$ and the other to $J = (N-2)/4 = 3/2$.
The next to the largest multiplet, with $j=N/2-1$, can be seen as the
largest multiplet corresponding to $N-2$ particles and one can
use the decomposition (\ref{NEVEN}) again. In the $N-2$ particles
case all the eigenvalues are nondegenerate while in the $N$ particle
case the same result is valid for $j=N/2-1$ but with some degeneracy for
the eigenvalues. In Table 2 we exhibit the multiplets $j$ and
their multiplicities $m_j$, the values of $J$ consistent with (\ref{NEVEN})
for each $j$ and the corresponding analytical solutions for the
eigenvalues obtained from the secular equation (\ref{SEC}).
One can check the consistency of these analytic expressions with those of 
Ref. \cite{LMG1} by using the
relation (\ref{V}) which gives $\delta$ in terms of $V/{\epsilon}$ \cite{CORR}.

In Fig. 2 we plot all positive eigenvalues of (\ref{HAM}) as a 
function of the strength $\delta$. One can notice an expected degeneracy
at $\delta = 0$ and some degeneracy at large values of $\delta$. In particular 
the largest eigenvalue with $j = 4, J = 2$ becomes degenerate
with the largest eigenvalue with $j = 3, J = 3/2$. 
The same eigenvalues with
opposite sign play the role of the ground state of the system,
which thus becomes degenerate for large value of $\delta$.

%%%%%%%%%%%%%%%%%%%%%%%%%%%%%%%%%%%%%%%%%%%%%%%%%
\section{Supplementary eigenvalues}
We have used the $sl(2,R)$ deformed algebra to study the spectrum of the 
Hamiltonian (\ref{HAM}). By construction, this algebra is richer than
the $su(2)$ algebra. Its representations have three labels 
instead of one, as for $su(2)$. Thus the number of representations is larger.
Moreover we can see that once $j$ and $c$
are fixed in (\ref{FG}), one can choose value of $J$ different from $j$  
such as the right hand side remains always positive. These values have
nothing to do with the LMG model.
However it turns out that some eigenvalues of Eq. (\ref{SEC})
associated to these representations are quite similar to those of 
(\ref{HAM}). It would be interesting to find out if they can be
related to a more general Hamiltonian than LMG. This is the subject of a 
further study.

%%%%%%%%%%%%%%%%%%%%%%%%%%%%%%%%%%%%%%%%%%%%%%%%%
\section{Summary}
Here we have presented a calculation of the whole spectrum of the
Lipkin-Meshkov-Glick Hamiltonian presented in the context of deformed
polynomial algebra. For any given number of particles $N$ the 
spectrum first divides into $j$ multiplets of the $su(2)$ algebra.
The eigenvalues associated with the largest $j$ are nondegenerate
except for $E = 0$. We have shown that the Hamiltonian
matrix of each $j$ further splits into two submatrices corresponding
to two distinct irreducible representations of the polynomial
deformed algebra. These representations bring new "quantum numbers",
one of them allowing us to distinguish between $N$ even and $N$ odd.
In order to illustrate the method 
we have derived explicit analytic expressions for the eigenvalues
of the LMG Hamiltonian for $N = 7$ and 8. Our method can be extended
to any $N$.

Furthermore, we have shown that the deformed polynomial algebra 
related to the LMG model implies a larger spectrum than that of
the model itself. Some of the new eigenvalues present characteristics
similar to those of the LMG model itself and may lead to a kind
of generalized model.

We hope this study could shed a new light into the LMG model and could
inspire further applications. \par

\vspace{1cm}
%%%%%%%%%%%%%%%%%%%%%%%%%%%%%%%%%%%%%%%%%%%%%%%%%%%%%%%%%%%%%%%%%%%%%%%%%
{\large\bf{Appendix }} \\
For the purpose of selfconsistency, here we recall a few results obtained
in Ref. \cite{DEBERGH} which are directly exploited in Sec. 3.
One of the aims of \cite{DEBERGH} was to construct finite dimensional
representations of the polynomial deformed algebra
\begin{equation}
[J_0,J_{\pm}] = \pm J_{\pm}
\end{equation}
\begin{equation}
[J_+,J_-] = \alpha J^3_0 + \beta J^2_0 + \gamma J_0 + \Delta~, 
~~~~~\alpha, \beta, \gamma, \Delta ~~\epsilon~~ \Re
\end{equation}
related to quasi-exactly soluble models.
Such representations imply the existence of kets, shortly denoted
by $|JM \rangle$, 
such that
\begin{eqnarray}\label{REPR}
J_0~ |JM \rangle = (M/q + c)~ |JM \rangle~,\nonumber \\
J_+~ |JM \rangle = f(M)~ |J,M + q \rangle~,\nonumber \\ 
J_-~ |JM \rangle = g(M)~ |J,M - q \rangle
\end{eqnarray}
In fact the kets are characterized by four labels: the number $J
= 0,\frac{1}{2},1,...$ related to an eigenvalue of the Casimir operator 
\begin{equation}
C = J_+ J_- + \frac{\alpha}{4}~ J^4_0 + (\frac{\beta}{3} - \frac{\alpha}{2})
~J^3_0 + (\frac{\alpha}{4} - \frac{\beta}{2} + \frac{\gamma}{2})~ J^2_0
+ (\frac{\beta}{6} - \frac{\gamma}{2} + \Delta)~ J_0
\end{equation}
which gives the dimension $2 J + 1$ of the representation, the number $M$
= -$J,-J+1,...,J$ which represents the eigenvalues of $J_0$, the positive
integer $q$ connected to the strength of the raising and lowering operators
$J_{\pm}$ and the shift real number $c$ which also enters the eigenvalues of
$C$. Thus $c$ depends on $J$.

The highest weight vectors of each representation impose the constraints
\begin{eqnarray}
f(J) = f(J-1) = ... = f(J-q+1) = 0~, \nonumber\\  
g(-J)= g(-J+1)= ... = g(-J+q-1) = 0~.
\end{eqnarray}
These relations ensure that the dimension of the representation
is $2J+1$ and
they lead to the following system of $q$ equations for the number $c$
\begin{eqnarray}\label{EQC}
\alpha~[ c^3 + \frac{3(d-l)}{2q}c^2 + \frac{J^2 - J(d + l) + 
l^2 - dl + d^2}{q^2}c + \frac{2J -d - l}{2q}c   \nonumber \\
+ \frac{2 J^2 (d-l) -2J (d^2 - l^2) + d^3 - d^2 l + d l^2 - l^3}{4 q^3}
%\nonumber \\
+ \frac{l^2 - d^2 + 2J(d-l)}{4 q^2} ] \nonumber \\
+ \beta~[ c^2 + \frac{(d-l)c}{q} + \frac{J^2 - J(d+l) + d^2 - dl + l^2 }{3 q^2} 
+ \frac{2J - d - l)}{6q}] \nonumber \\
+ \gamma~(c + \frac{d-l}{2q}) + \Delta &=& 0 
\end{eqnarray}
with $l = 0,1,...,q-1$. The non-negative integer $d$, introduced above for
convenience, has to take specific values according to $J$ and $l$. These
values have been given in \cite{DEBERGH}. 
For the real value of $\alpha, \beta, \gamma $ and
$\Delta$ of Eqs. (\ref{PARAM})
the obtained system of equations
is compatible if and only if $q = 1$ or $q = 2$. 

For $q = 1$ one has
$d = 0$ 
and one remains with a single equation
\begin{equation}\label{Q1}
\alpha c~ [c^2 + J(J + 1)] + \beta~ [c^2 + \frac{J(J + 1)}{3}] 
+ \gamma c + \Delta = 0.
\end{equation}
For $q = 2$, one has:  1) $d = 0$ when $l = 0$ and $J$ is an integer or   
when $l=1$ and $J$ is a half integer, 
2) $d = 1$ when $l = 0$ and $J$ is a half integer or when $l=1$ and $J$ is an 
integer, with corresponding equations of type (\ref{EQC}). 

The representations (\ref{REPR}) are completely specified with
\begin{eqnarray}
f(J-kq-q-l)g(J-kq-l) &=& (k+1) \nonumber \\
& & \{ \alpha~(\frac{J-l}{q}+c)^3 + 
\beta ~(\frac{J-l}{q}+c)^2 + \gamma~(\frac{J-l}{q}+c) + \Delta \nonumber \\
& & -\frac{1}{2} [3 \alpha~(\frac{J-l}{q}+c)^2 + 2 \beta~(\frac{J-l}{q}+c)
+ \gamma ]~k  \nonumber \\
& & + \frac{1}{6} [ 3 \alpha~(\frac{J-l}{q}+c) + \beta ]~k(2 k + 1) \nonumber \\
& & - \frac{\alpha}{4} ~k^2 (k + 1)  \}
\end{eqnarray}
where $l = 0,1,...,q-1$ and $k = 0,1,...,(2J-d-l)/q$. \par

%%%%%%%%%%%%%%%%%%%%%%%%%%%%%%%%%%%%%%%%%%%%%%%%%%%%%%%%%%%%%%%%%%%%%%%%
\vspace{2cm}
{\bf Acknowledgements}. 
One of us (F.S.) is most grateful for warm hospitality
and living support at ECT*, Trento.  
%%%%%%%%%%%%%%%%%%%%%%%%%%%%%%%%%%%%%%%%%%%%%%%%%%%%%%%%%%%%%%%%%%%%%%%%%

%\newpage
%%%%%%%%%%%%%%%%%%%%%%%%%%%%%%%%%%%%%%%%%%%%%%%%%%%%%%%%%%%%%%%%%%%%%%%%%
%\listoftables
\listoffigures
%%%%%%%%%%%%%%%%%%%%%%%%%%%%%%%%%%%%%%%%%%%%%%%%%%%%%%%%%%%%%%%%%%%%%%%%%
\begin{table}\label{N7}
\caption[]{Eigenvalues of the Hamiltonian (\ref{HAM})
for $N = 7$ particles.}
\begin{center}
\begin{tabular}{|c|c|c|l|}
%\hline
$j$ & $m_j$ & $J$  & $Eigenvalues$ \\
\hline
~1/2~ & 14~ & 0 & $\pm$ 1/2 \\
~3/2~ & 14~ & 1/2 & $\pm (~~\frac{1}{2} \pm \sqrt{1+ \frac{3}{49} 
{\delta}^2}~~)$~~~~~ \\
~5/2~ & 6~  &  1 & \\
~7/2~ & 1~  & 3/2 & 
\end{tabular}
\end{center}
\end{table}
%%%%%%%%%%%%%%%%%%%%%%%%%%%%%%%%%%%%%%%%%%%%%%%%%%%%%%%%%%%%%%%%%%%%%%%%
\begin{table}\label{N8}
\caption[]{Eigenvalues of the Hamiltonian (\ref{HAM})
for $N = 8$ particles.}
\begin{center}
\begin{tabular}{|c|c|c|l|}
%\hline
$j$ & $m_j$ & $J$  & $Eigenvalues$ \\
\hline
~0~ & 14~ & 0 & 0 \\
~1~ & 28~ & 0 & 0 \\ 
~~ & ~ & 1/2 & ${\pm \sqrt{1+\frac{1}{64} {\delta}^2}}$~~ \\
~2~ & 20~ & 1/2 & ${\pm \sqrt{1+\frac{9}{64} {\delta}^2}}$~~ \\
~~ & ~ & 1   & 0,~~~${\pm \sqrt{4+\frac{3}{16} {\delta}^2}}$~~ \\
~3~ &  7~ & 1   & 0,~~~${\pm \sqrt{4+\frac{15}{16} {\delta}^2}}$~~ \\
~~ &  ~ & 3/2 & ${\pm \sqrt{5+\frac{33}{64} {\delta}^2 
\pm \sqrt{16+ \frac{3}{2} {\delta}^2 + \frac{27}{128}} {\delta}^4}}$~~ \\
~4~ &  1~ & 3/2 & ${\pm \sqrt{5+\frac{113}{64} {\delta}^2
\pm \sqrt{16+ \frac{19}{2} {\delta}^2 + \frac{275}{128}} {\delta}^4}}$~~ \\
~~ &  ~ & 2   & 0,~~~${\pm \sqrt{10+\frac{59}{32} {\delta}^2
\pm \sqrt{36 - \frac{9}{8} {\delta}^2 + \frac{2025}{1024}} {\delta}^4}}$~~ 
\end{tabular}
\end{center}
\end{table}

%%%%%%%%%%%%%%%%%%%%%%%%%%%%%%%%%%%%%%%%%%%%%%%%%%%%%%%%%%%%%%%%%%%%%%%%%%%
\newpage
\begin{figure}
\begin{center}
\psfig{figure=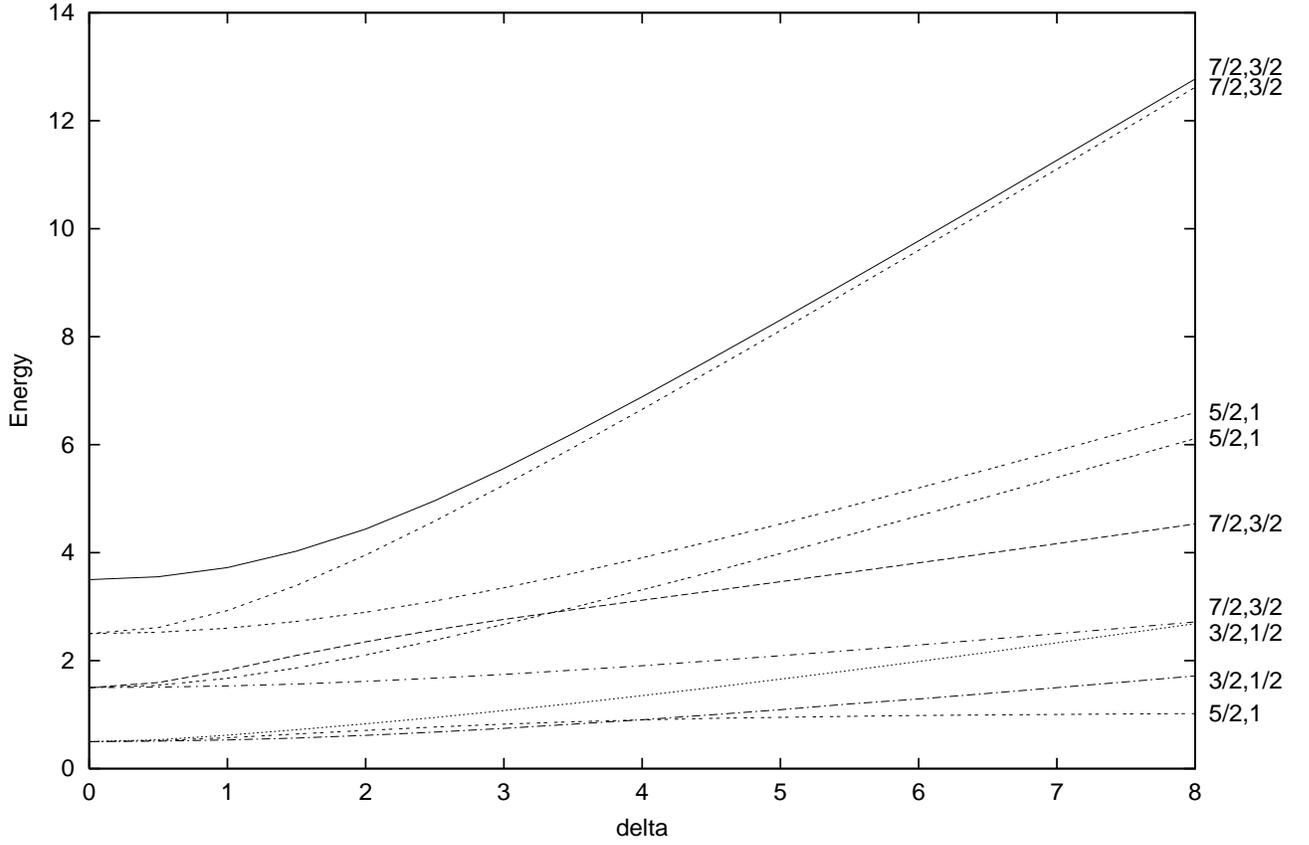,width=16.5cm}
\end{center}
\caption{\label{Fig1}Positive eigenvalues of the LMG Hamiltonian (\ref{HAM}),
as a function of the parameter $\delta$ defined by Eq. (\ref{V})
for $N = 7$. The eigenvalues are labelled
by $j,J$ and correspond to the rows 3 and 4 of Table 1.}  
\end{figure}

\newpage
\begin{figure}
\begin{center}
\psfig{figure=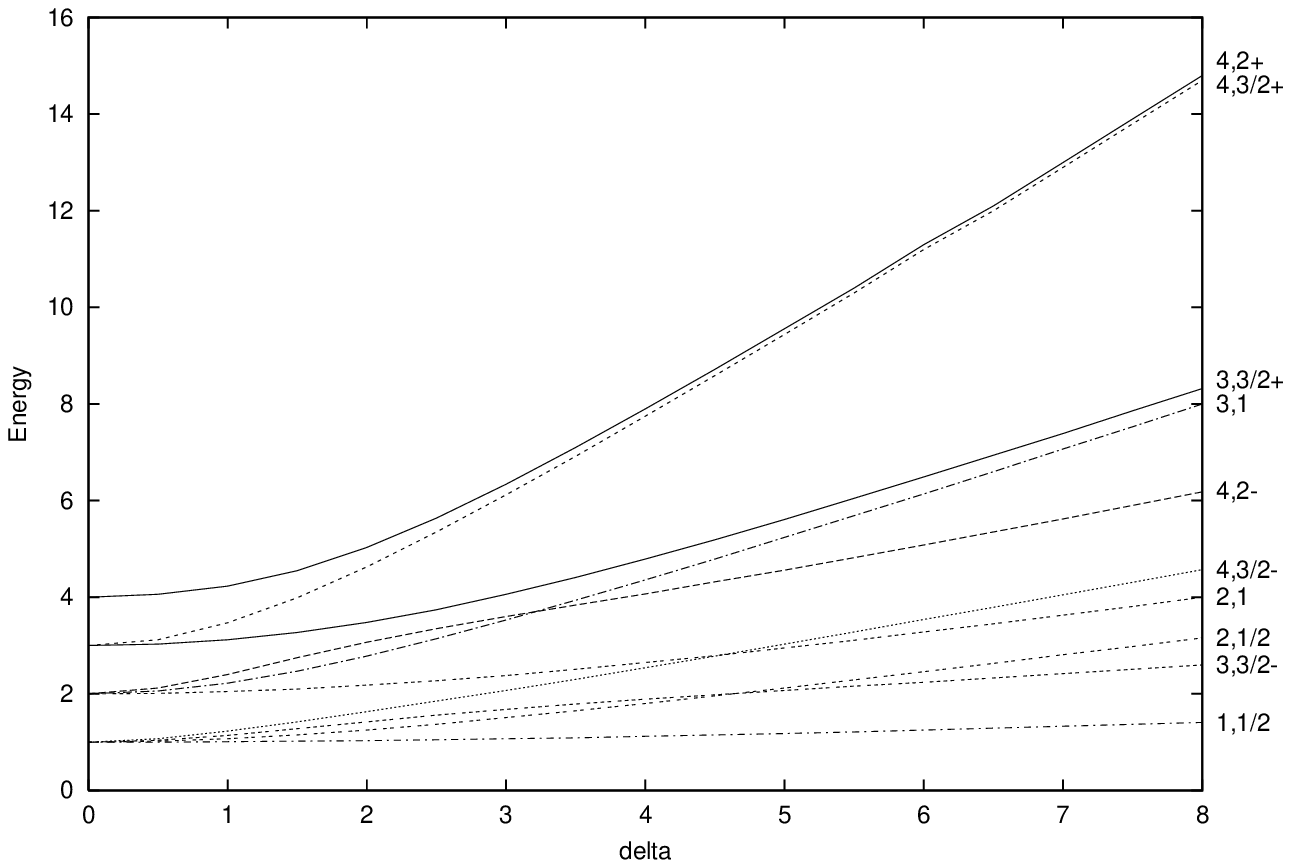,width=16.5cm}
\end{center}
\caption{\label{Fig2} Positive eigenvalues of the LMG Hamiltonian (\ref{HAM}),
as a function of the parameter $\delta$ defined by Eq. (\ref{V})
for $N = 8$. The eigenvalues are labelled either by $j,J$ or by $j,J,sign$
when necessary, where $sign$ means the sign in front of the inner
square root in the last column of Table 2. } 
\end{figure}


\begin{references}
\bibitem{CORR} Note that in Ref. \cite{LMG1} above there are some printing 
mistakes.
In the first row of Eqs. (3.5) a factor of 4 is missing in front of
the inner square root and in Eqs. (3.5) a factor of 6 is also missing
in front of the inner square root.
\end{references}
\end{document}